\documentclass[11pt]{amsart} 

\usepackage{amsmath}  % needed for \tfrac, \bmatrix, etc.
\usepackage{amsfonts} % needed for bold Greek, Fraktur, and blackboard bold
\usepackage{graphicx} % needed for figures
\usepackage{bm}
\usepackage[svgnames,dvipsnames]{xcolor}

\begin{document}

\title[]{Some problem with the Lorentz transformations of energy and momentum of the electromagnetic field}

\author{Vladimir Onoochin}
%\email{onoochin@gmail.com}
%\affiliation{Sirius, LLC, 3 Nikoloyamsky lane, Moscow, Russia}

\begin{abstract}
In this work, it is shown that the energy and momentum of electromagnetic fields created by a classical charge, whose velocity varies with time, do not form four-vector. 

A possible explanation for this result is that the calculation of energy and momentum is performed as an integration of the densities of these quantities, {\it i.e.}  the squares of the electromagnetic field $E^2$ and $H^2$ in the whole space at the hyperplane $t=const$. But $E^2$ and $H^2$, which are calculated at a fixed point in time, $t$, are all created at previous (retarded) instants of time. In other words, all densities $E^2$ and $H^2$ are independent of each other. Meanwhile, Lorentz transformations are defined as transformations between physically connected quantities.
\end{abstract}

\maketitle

\section{Introduction}

The total four-momentum of a charged system can be written as~\cite{Iv}
\begin{equation}
P^{\mu}_{tot}=\frac{1}{c}\int \left[\Theta^{\mu\,\nu}+{\rm T}^{\mu\,\nu}\right] d  ^3x\,,\label{1}
\end{equation}
where $\Theta^{\mu\,\nu}$ is a stress-energy tensor which describes non-electromagnetic forces and matter while ${\rm T}^{\mu\,\nu}$ is the energy-momentum density tensor of the electromagnetic field. The integral is calculated over the hyperplane $t=const$.

If the charged system is represented by the only classical electron, the integral of $\Theta^{\mu\,\nu}$ yields its mechanical energy and momentum which are components of the four--vector. Therefore, the integral of ${\rm T}^{\mu\,\nu}$ represents the four-vector of the energy and momentum of the electromagnetic field created by this electron,
\begin{eqnarray}
P^0_{field}=\mathcal{E}=\int \frac{E^2+H^2}{8\pi}d  V\,;\\
P^i_{field}={\bf P}=\int \frac{\left[ {\bf E}\times {\bf H}\right]}{4\pi c} d  V\,,\,\,i=1,2,3\,,
\end{eqnarray}
where ${\bf E}$ and ${\bf H}$ are the electric and magnetic fields created by this charge in vacuum. 

As any four-vector, $P^{\mu}_{field}$ can be determined in another inertial frame K$^\prime$, moving relatively to original frame K with constant velocity $v$. So the zero component of this vector in K$^\prime$ is found as 
\begin{equation}
P^0_{field}=\mathcal{E}'=\frac{1}{\sqrt{1-(v/c)^2}}\left[\mathcal{E}-{\bf v}\cdot {\bf P}\right]\,, \label{rel}
\end{equation}
where $\mathcal{E}$ and ${\bf P}$ are the electromagnetic energy and momentum of the charge in K.

It should be noted that  only a few number of examples of the above relation verification are given in scientific literature. Griffiths notes that  since ``electromagnetic stress-energy tensor is not by itself divergenceless,...electromagnetic energy and momentum do not constitute a four-vector''~\cite{Gr}.  However, as it will be shown below by direct calculations, the energy and momentum of a uniformly moving classical charge are components of a four-vector which transforms in accordance to the Lorentz transformations.

Boyer~\cite{Bo} presents several examples of  electromagnetic energy and momentum calculations for systems of  moving charges. But the calculations of this author are made in the Darwin approximation and, therefore, cannot be considered as a rigorous proof of the loss of the relativistic properties for these physical quantities. Moreover, rigorous mathematical calculations of the electromagnetic energy and momentum generated by a classical charge are not available in the scientific literature. This is partly due to the difficulties in calculating integrals containing expressions for EM fields. But basically the absence of such examples is due to the generally accepted opinion that the energy and momentum of all physical objects must form a four-vector. Therefore, adding such examples to the text of classical electrodynamics courses would be superfluous.

Meanwhile it could be useful to calculate energy and momentum of the EM field created by a classical electron in two inertial frames. Such an example can be used to demonstrate how the Lorentz transformation (LT) of certain physical quantity is established. 

Since the field energy and momentum are represented by integrals over the whole space, expressions for electromagnetic fields should be given in the present time variables because integrals of retarded quantities cannot be computed. In classical electrodynamics, only two sets of expressions for the EM fields are given in these variables, {\it i.e.} when the classical electron, that creates the required fields, moves uniformly or with constant acceleration.

Corresponding calculations will be made in Sections II and III. If in the first case, {\it i.e.} for a uniform motion of the classical electron the electromagnetic energy and momentum of its field form a four-vector, it is not the case for a uniformly accelerated motion of this charge. Some explanation of the absence of relativistic properties of the electromagnetic energy and momentum will be given in the final section of this work.

\section{Electromagnetic energy and momentum of the classical electron being in uniform motion.}

Let us show that the energy and momentum of the EM field created by a uniformly moving charge form the components of a four-vector. To do this, it is necessary to calculate these quantities in two inertial frames, for example, in frame K$_0$, where the charge is at rest (comoving frame), and in frame K, moving at a constant speed $v$ relative to K$_0. $.

The energy of the electric field created by classical electron being at rest is 
\begin{equation}
\mathcal{E}_0=\int \frac{E^2_{v=0} }{8\pi}d  V_0 =\int \limits_{R_0}^{\infty}\frac{e^2}{2r_0^4}r_0^2d  r_0 =\frac{e^2}{2R_0}\,,
\end{equation}
where $r_0=\sqrt{x_0^2+y_0^2+z_0^2}$ and $R_0$ is classical radius of the electron. The index \lq{}0\rq{} means that the quantities are defined in the K$_0$ frame. The above integral can be written as
\begin{equation}
\int \frac{E^2_{v=0} }{8\pi}d  V_0 =\int \frac{E_{\|}^2+E_{\bot}^2 }{8\pi}d  V_0 \,,
\end{equation}
where $E_{\|}$ is the longitudinal (along the axis of motion of the electron) component; it is assumed that the directions of the $x$ and $x_0$ coincide. For the charge at rest, $E_{\|}=qx_0/r_0^3$ and the transverse component of the electric field is 
$E_{\bot}=q\sqrt{y_0^2+z_0^2}/r_0^3$.

The electric field of a uniformly moving charge is
\begin{equation}
{\bf E}=\frac{q}{\sqrt{1-(v/c)^2}}	\dfrac{{\bf r}}{\left[\dfrac{(x-vt)^2}{1-(v/c)^2}+y^2+z^2\right]^{3/2}} \,,
\end{equation}	
where ${\bf r}=\{(x-vt);y;z \}$. The square of the electric field is
\begin{equation}
E^2=\frac{q^2}{1-(v/c)^2}	\dfrac{(x-vt)^2+y^2+z^2}{\left[\dfrac{(x-vt)^2}{1-(v/c)^2}+y^2+z^2\right]^{3}} \,.
\end{equation}	
Let us introduce spherical coordinates,
\begin{eqnarray*}
R=\sqrt{\dfrac{(x-vt)^2}{1-(v/c)^2}+y^2+z^2}\,\,;\,\, y^2+z^2=R^2\sin^2\theta\,\,;\\
 d  xd  yd  z=\sqrt{1-(v/c)^2}d  x'd  yd  z=\sqrt{1-(v/c)^2} R^2d  R \sin\theta d \theta d \phi\,.
\end{eqnarray*}
Then 
\begin{equation}
E^2=\frac{q^2}{1-(v/c)^2}	\dfrac{\left[1-(v/c)^2\right] \cos^2\theta+\sin^2\theta}{R^4} \,.
\end{equation}
It can be rewritten as
\begin{equation}
E^2=\frac{1}{1-(v/c)^2}\left\{\left[1-(v/c)^2\right]E_{\|}^2+E_{\bot}^2\right\}\,.
\end{equation}
The relation between the transverse $E_{\bot}(v)$ field and circular $H_{\phi}(v)$ field in K frame is (Eq.~(11.150) of~\cite{JDJ}),
\begin{equation}
H_{\phi}(v)=\frac{v}{c}E_{\bot}(v)\,. \label{cond}
\end{equation}
Therefore, the energy density of both fields is
\begin{equation}
\varepsilon =\frac{1}{1-(v/c)^2}\frac{E^2+H^2}{8\pi}=\frac{1}{1-(v/c)^2}\frac{\left[ (1-v^2/c^2)E_{\|}^2+(1+v^2/c^2)E_{\bot}^2\right] }{8\pi}\,.
\end{equation}
Now let  us verify relation~\eqref{rel} where 
\begin{equation}
\mathcal{E}_v=\int \varepsilon  d  V'=\sqrt{1-(v/c)^2}\int 	 \varepsilon d  x' d  yd  z \, ,
\end{equation}
and 
\begin{eqnarray}
{\bf P}=\frac{\sqrt{1-(v/c)^2}}{4\pi}\int {\bf p}d  V' =\frac{1}{4\pi c\sqrt{1-(v/c)^2}}\int [E_{\bot}\times H_{\phi}]  d  x' d  yd  z=\nonumber\\
=\frac{1}{8\pi c^2\sqrt{1-(v/c)^2}}\int 2{\bf v}E_{\bot}^2 d  x' d  yd  z  \, ,
\end{eqnarray}
where ${\bf p}$  is the EM momentum density.

So Eq.~\eqref{rel} becomes
\begin{eqnarray}
&&\mathcal{E}_0=\int\frac{E_{\|}^2+E_{\bot}^2}{8\pi} d  x_0d  y_0d  z_0\,;\label{0}\\
&&\frac{\mathcal{E}_v-\left({\bf v}\cdot{\bf P} \right)}{\sqrt{1-(v/c)^2}}=
\frac{1}{1-(v/c)^2}\int \frac{\left[ (1-v^2/c^2)E_{\|}^2+(1-v^2/c^2)E_{\bot}^2\right] }{8\pi} d  x' d  yd  z =\nonumber\\
&&=\int \frac{\left[ E_{\|}^2+E_{\bot}^2\right] }{8\pi} d  x' d  yd  z \,   .\label{v}
\end{eqnarray}
%The lower limit in~\eqref{v} is radius of the classical electron. For moving electron, its shape changes from spherical to the shpe of contracted ellipsoid. It seems that the radius of the moving electron is different in the $x$ and $y,\,z$ directions. But this change is compensated by introduction of the factor $\sqrt{1-(v/c)^2}$ in volume element of integration. In the other words, in the coordinates $x'=(x-vt)/\sqrt{1-(v/c)^2}$, $y,\,z$ the shape of the moving electron is spherical.

Thus it is shown that calculation of integral~\eqref{0}, written in the coordinates $x_0,y_0,z_0$, gives the same value that calculation of integral~\eqref{v}, written in the coordinates $x',y,z$. It means that the LT are performed for the energy and momentum of the EM field created by a uniformly moving charge.

Similar relation is given in the textbook of Jackson~\cite{JDJ}, who, following Rohrlich~\cite{Rohr}, gives another definition of the zeroth component of the four-momentum,
\begin{equation}
	P^0_{Rohr}=\frac{1}{\sqrt{1-(v/c)^2}}\int\left[\mathcal{E}-({\bf v}\cdot{\bf P})\right]d  ^3 r  \,  .
\end{equation}	
According to these authors, this zeroth component should be invariant in any inertial frame:
\begin{equation}
\varepsilon-({\bf v}\cdot{\bf p})=\frac{E^2+H^2}{8\pi}-
\frac{\left({\bf v}\cdot [{\bf E}\times{\bf H}]\right)}{4\pi c}=
\frac{E^2-H^2}{8\pi}\,, \label{mass}
\end{equation}
where Jackson uses ${\bf B}=  [{\bf v}\times{\bf E}]/c$. Because $E^2-H^2$ is invariant, the relativistic rule
\begin{equation}
\int \frac{E_{v=0}^2}{8\pi} dV_0 = \frac{1}{\sqrt{1-(v/c)^2}} \int \frac{E^2-H^2}{8\pi} dV \,,\label{M}
\end{equation}
should be fulfilled in the general case. However, it is not so. It will be shown in the next section that the electromagnetic energy and momentum of a uniformly accelerated classical electron deviate from  this rule. 

\section{Electromagnetic energy and momentum of the classical electron being in uniformly accelerated motion}

In this section the calculation procedure developed in Section II will be repeated for the fields created by a uniformly accelerated charge. The essential property of these fields is that they are expressed in present time variables, and the integration of the energy and momentum densities  does not seem to have obstacles.

It should be especially noted that uniformly accelerated motion of a classical charge can be realized only if a constant external electric field is applied to the system~\cite{Sch}. Formally, this constant electric field should be involved into relativistic transformation of the total energy and momentum.
However, as was shown by Born~\cite{Born},  the mechanical energy and momentum of this charge form the four-vector without taking into account  the external field. In other words, a uniformly accelerated motion of a classical charge can be initially given and the mechanical $\mathcal{E}_{\rm mech}$ and  ${\bf P}_{\rm mech}$ are the components of a four-vector. Since the total energy and momentum of the system  must be four-vector, and the first term in the integral~(\ref{1}) is the four-vector, the energy and momentum of the EM fields created by such a charge must also be four-vector.

The expressions for EM fields of a uniformly accelerated charge are  given by  Eq. (85) of~\cite{Sch},  
\begin{equation}
E_{\|}=E_x=\frac{4k^2(x^2-\omega ^2-\xi ^2)}{\left [ (x^2+\omega ^2-\xi ^2)^2+4k^2\omega ^2\right]^{3/2}}\,;\quad
E_{\bot}=E_{\omega}=\frac{8k^2 x\omega }{\left [ (x^2+\omega ^2-\xi ^2)^2+4k^2\omega ^2\right]^{3/2}}\,,\label{Esh}
\end{equation}
\begin{equation}
H_{\phi}=\frac{8k^2 ct\omega }{\left [ (x^2+\omega ^2-\xi ^2)^2+4k^2\omega ^2\right]^{3/2}}\,, \label{Hsh}
\end{equation}
where $\omega=\sqrt{y^2+z^2}$ in notation of~\cite{Sch}, and 
\[
\xi (t)=\sqrt{k^2+c^2t^2}\,;\quad v(t)=\frac{c^2t}{ \sqrt{k^2+c^2t^2}}\,;\quad \frac{1}{\sqrt{1-(v/c)^2}}=
\frac{\sqrt{k^2+c^2t^2}}{k}\,.
\] 
Strictly speaking, Eqs.~\eqref{Esh} and~\eqref{Hsh} contain the factor in a form of the step function $\theta(x+ct)$ which describes propagation of the wavefront of the electric field. This factor will be used to determine an integration area of the EM fields.

The {\it lhs} of Eq.~\eqref{rel} is calculated in the K$_0$ frame where the charge is assumed to be at rest. For the hyperbolic motion of a classical charge, the definition of such a frame is given by Pauli on p. 93 in Sec. 32($\gamma$) of~\cite{Pau} .
 Corresponding expressions for the EM fields of this charge  have the form(Eqs. (250) of the cited textbook),  
\begin{equation}
E_{\|}(0)=\frac{4k^2(x_0^2-\omega_0 ^2-k ^2)}{\left [ (x_0^2+\omega_0 ^2-k ^2)^2+4k^2\omega_0 ^2\right]^{3/2}}\,;\quad
E_{\bot}(0)=\frac{8k^2 x_0\omega_0 }{\left [ (x_0^2+\omega_0 ^2-k ^2)^2+4k^2\omega_0 ^2\right]^{3/2}}\,, \label{Esh0}
\end{equation}
\[
H_{\phi}(0)=0  \,  .
\]
where $x_0$ and $\omega_0$ are the coordinates of detection of the ${\bf E}$ field when the charge is located at $\xi_{t=0}=k$ in K$_0$.

The above expressions can be used to verify Eq.~(\ref{rel}). But before -- to simplify calculations -- let us eliminate the magnetic field from the energy and momentum densities by means of~\eqref{Esh} and~\eqref{Hsh}.  Then  Eq.~\eqref{rel} for Schott's fields becomes,
\begin{equation}
\int\left[ \frac{E_{\|}^2+E_{\bot}^2 }{8\pi}\right]_{v=0}d  V_0 =\frac{1}{\sqrt{1-v^2}}
\int\left[ \frac{E_{\|}^2+(x-ct)^2E_{\bot}^2 /x^2}{8\pi}\right]_{v\neq 0}d  V \,, \label{S-rel}
\end{equation}
where the components of the electric field are given by Eqs.~\eqref{Esh0} and~\eqref{Esh}, or
\begin{equation}
\mathcal{E}_0=\frac{2k^4}{\pi}\int\limits_0^{2\pi}d  \phi \int\limits_0^{\infty}\omega_0d  \omega_0 \int \limits_0^{\infty}d  x_0\frac{ (x_0^2-\omega_0 ^2-k ^2)^2+ 4 x_0^2\omega_0^2}{\left [ (x_0^2+\omega _0^2-k ^2)^2+4k^2\omega_0 ^2\right] ^3}\,. \label{E20}
\end{equation}
\begin{equation}
\frac{\mathcal{E}_v-\left({\bf v}\cdot{\bf P} \right)}{\sqrt{1-(v/c)^2}}=\frac{2k^3\sqrt{k^2+c^2t^2}}{\pi}\int\limits_0^{2\pi}d  \phi \int\limits_0^{\infty}\omega d  \omega \int \limits_{-ct}^{\infty}d  x\frac{ (x^2-\omega ^2-k ^2-c^2t^2)^2+
4 (x-ct)^2\omega^2}{\left [ (x^2+\omega ^2-k^2-c^2t^2)^2+4k^2\omega ^2\right]^3}\,. \label{E2v}
\end{equation}
where the limits of integration over the $x_0$ and $x$ variables are determined by presence of $\theta$--function. 

Integral~\eqref{E2v} cannot be evaluated in closed form. So, to show that the values of the integrals~\eqref{E20} and~\eqref{E2v} are not equal to each other, we use the condition according to which, in accordance with the special theory of relativity, ~\eqref{rel} must hold for any $t$ . 

Let us rewrite the condition~\eqref{rel} in a form
\begin{equation}
\mathcal{E}_0=\frac{\mathcal{E}_v-\left({\bf v}\cdot{\bf P}\right)}{\sqrt{1-(v/c)^2}}\,\,\to\,\,
\mathcal{E}_0-\frac{\mathcal{E}_v-\left({\bf v}\cdot{\bf P}\right)}{\sqrt{1-(v/c)^2}}=0\,, \label{r}
\end{equation}
If ~\eqref{r}  holds for any $t$ , let us expand its {\it lhs}  into a series in $t$ and, after simplification, calculate the integrals. But first it is needed to shift the lower limit of $x$ to 0 using $x+ct=x'$. Then, since the regions of integration in~\eqref{E20} and~\eqref{E2v} are identical, it is possible to perform the integrand. After changing $x+ct=x'\to x$, $\omega_0\to\omega$, $x_0\to x$ and making integration with respect to $\phi$,  we have for this difference between the integrands,
\begin{eqnarray}\nonumber
4k^4\,\cdot\frac{ (x^2-\omega ^2-k ^2)^2+ 4 x^2\omega^2}{\left [ (x^2+\omega^2-k ^2)^2+4k^2\omega ^2\right] ^3}& - &4k^3\sqrt{k^2+c^2t^2}\cdot\frac{ (x(x-2ct)-\omega ^2-k ^2)^2+ 4 (x-2ct)^2\omega^2}
{\left [ (x(x-2ct)+\omega^2-k ^2)^2+4k^2\omega ^2\right] ^3}\approx\\
&\approx &-\frac{32 k^4  x (k^2 - x^2)}{\left[k^4 + 2 k^2 (\omega^2 - x^2) + 
(\omega^2 + x^2)^2\right]^3}\, t + \mathcal{O}[t^2]\,.
\end{eqnarray}
The term of expansion that is proportional to $t$ has a singularity at $x\,\to\,k$, $\omega\,\to\,0$ which should be removed by introducing the classical radius of the electron, $R_0=\sqrt{x_r^2+\omega_r^2}$.

To determine whether the energy and momentum of Schott's electromagnetic field obey the condition~\eqref{r}, it is necessary to calculate the integral of 
\[
{\tt Int}=-\frac{32 k^4  x (k^2 - x^2)}{\left[k^4 + 2 k^2 (\omega^2 - x^2) + (\omega^2 + x^2)^2\right]^3}
\]
over all space ($x,\,\omega$) where the lower limit should be equal to the classical radius of electron.  Introduction of this parameter, {\it i.e.} the classical radius, transforms the improper integral of {\tt Int} into the integral of proper type, which value does not depend on the order of integration.

Because this integrand changes its sign at $x=k$, it is possible to compute the integral over the $x$ variable,
\begin{eqnarray}\nonumber
F(k,\omega)& =& \int\limits_0^{\infty}{\tt Int}(x,k,\omega)d  x= -\frac{3\pi k^8+ 12k^7\omega +12\pi k^6\omega^2 -84k^5\omega^3
+ 18\pi k^4\omega^4 - 44k^3\omega^5}{ 8 k \omega^3 (k^2 + \omega^2)^4}-\\
&-&\frac{12\pi k^2\omega^6 -12k\omega^7 + 3\pi \omega^8+6 (k^2 + \omega^2)^4 \arctan[k/(2 \omega) - \omega/(2 k)]}
{ { 16 k \omega^3 (k^2 + \omega^2)^4}}\,. \label{w}
\end{eqnarray}
The next step is integration of $F(k,\omega)$ over $\omega $ in the limits $\omega \in[R_0;\infty]$. Despite the fact that the exact value of the classical electron radius is unknown, this is not an obstacle to obtaining the needed result. Numerical computation of the integral for values of the parameters $R_0=0.01;\,k=1$ gives the value of the {\it lhs} of Eq.~\eqref{r} as  $\approx -2943t$. For $R_0=0.01;\,k=0.1$ this value is $\approx -27690t$., and for  $R_0=0.005;\,k=0.1$ this value is $\approx -115850t$, (multiplied by 23.04$\cdot10^{-20}$ esu$^2$/(cm$\cdot$ sec); the calculations are made in Gaussian units). This allows us to conclude: the non-zero difference in energy-momentum calculated in two inertial frames shows that these electrodynamic quantities do not have relativistic properties in the considered case.

\section{Conclusions}

In this work, the electromagnetic energy and momentum of a classical electron are calculated in two cases, when the electron moves uniformly and when is moves with constant acceleration. According to the special relativity, these quantities calculated in two inertial frames should be connected by the Lorentz transformations~\eqref{rel}  as components of a four-vector. But if for an electron moving uniformly  the relation between $\mathcal{E}_0$ and  ($\mathcal{E}_v,\,{\bf P}$) holds, it is not the case for the electron being in hyperbolic motion. 

Because two different results are obtained, {\it i.e.} the fulfillment of~\eqref{rel} in the first case and violation of~\eqref{rel} in the second case,  it would be reasonable to explain why this difference appears.

From the author's point of view, the difference in the results arises because the integrals representing energy and momentum have to be calculated at the hyperplane $t=const$. 

When a classical electron moves uniformly, its behavior can be reduced to the stationary case. By the appropriate transformation of coordinates, a uniformly moving electron is represented as a charge at rest, whose EM field is static~\cite{PP}. In this case, the instant of creation of an electric field, detected at some remote point, can be any. It can be assumed that the field is created instantly at all points in space. Moreover, this field, like its energy, does not change with time. So choice of the hyperplane is not essential. 
 
But in the case of a uniformly accelerated motion of a classical electron, the reduction of the problem to the stationary case is impossible. Schott derived expressions~\eqref{Esh} and~\eqref{Hsh} from the Lienard--Wiechert potentials, according to which the electromagnetic fields detected at some point in space at time $t^\star$ are emitted by the charge at the previous (retarded) time $t_r<t^\star $.  Thus, the EM fields detected at some point $\mathcal{P}(x,y,z)$ in space at $t^\star$ are not the fields created by the charge located at the point $\mathcal{O}(x_{ch},0,0)$ at the same time $t^\star$. It means that there is no causal connection between the charge being at some instant $t^\star$ at some point $\mathcal{O}$ and the fields detected at the same instant but at different points $\mathcal{P}$ of space. So the energy and momentum densities of every amounts of the EM field located at some element of volume are the set of independent quantities, and integration of these quantities over the whole space, made in the K$_0$ frame for $t=0$, is a formal summation. Similar integration made in the K frame for the given $t^\star$ is also the formal summation. There is no physical factor connecting the results of these summations. The Lorentz transformations always establish some physical connection between the transformed quantities. If the physical connection between two parameters calculated in two inertial frames is absent, the LT between these parameters cannot be established.

The author does not state the aim to verify whether  the energy and momentum form the four-vector or not for the EM field, created by a classical charge being in arbitrary motion. But from the other side, if $\mathcal{E}$ and ${\bf P}$  do not form the four-vector for the fields of the charge being in hyperbolic motion, one should be careful to apply the Lorentz transformation of these quantities without corresponding verification of their correctness in the same way as it is done in Sec. III of this work. 

%\subsection*{Conflict of interest}

%The author has no conflicts to disclose.

\end{document}